\documentclass[lineno]{biometrika}

\usepackage{amsmath}

\usepackage{times}
\usepackage{bm}
\usepackage{natbib}

\usepackage[plain,noend]{algorithm2e}

\makeatletter
\renewcommand{\algocf@captiontext}[2]{#1\algocf@typo. \AlCapFnt{}#2} 
\def\@algocf@capt@plain{top}
\renewcommand{\algocf@makecaption}[2]{%
  \addtolength{\hsize}{\algomargin}%
  \sbox\@tempboxa{\algocf@captiontext{#1}{#2}}%
  \ifdim\wd\@tempboxa >\hsize
    \hskip .5\algomargin%
    \parbox[t]{\hsize}{\algocf@captiontext{#1}{#2}}
  \else%
    \global\@minipagefalse%
    \hbox to\hsize{\box\@tempboxa}
  \fi%
  \addtolength{\hsize}{-\algomargin}%
}
\makeatother

\usepackage{amsbsy}
\usepackage{mathrsfs}
\usepackage{amssymb}
\usepackage{graphicx}
\usepackage{array,booktabs}
\usepackage{multirow}
\usepackage{caption}
\usepackage{subcaption}

\newcommand{\like}{\log{P}}

\newcommand{\Poi}{\text{Poi}}
\newcommand{\Exp}[1]{E\!\left( #1 \right)}
\newcommand{\Var}[1]{\mathrm{var}\!\left( #1 \right)}  
\newcommand{\Cov}[1]{\mathrm{cov}\!\left( #1 \right)}

\renewcommand{\mid}{\,|\,}
\newcommand{\e}{e}

\newcommand{\Dbar}{\overline{D}}

\begin{document}


\jname{Biometrika}
\jyear{---}
\jvol{---}
\jnum{---}
\markboth{Yan {\em et al.}}{Model Selection for Degree-corrected Block Models}

\title{Model Selection for Degree-corrected Block Models}

\author{Xiaoran Yan}
\affil{Computer Science, University of New Mexico \email{everyan@cs.unm.edu}}
\author{Cosma Rohilla Shalizi}
\affil{Statistics Department, Carnegie Mellon University, \email{cshalizi@cmu.edu} }
\author{Jacob E. Jensen}
\affil{Computer Science, Columbia University, \email{2timesjay@gmail.com}}
\author{Florent Krzakala}
\affil{ESPCI ParisTech and CNRS, \email{fk@espci.fr}}
\author{Cristopher Moore}
\affil{Santa Fe Institute and University of New Mexico, \email{moore@santafe.edu} }
\author{Lenka Zdeborov\'a}
\affil{Institut de Physique Th\'eorique, CEA Saclay and CNRS, \email{lenka.zdeborova@cea.fr}}
\author{Pan Zhang}
\affil{ESPCI ParisTech and CNRS, \email{pan.zhang@espci.fr}}
\author{Yaojia Zhu}
\affil{Computer Science, University of New Mexico, \email{yaojia.zhu@gmail.com} }

\maketitle

\begin{abstract}
  The proliferation of models for networks raises challenging problems of model
  selection: the data are sparse and globally dependent, and models are
  typically high-dimensional and have large numbers of latent variables.
  Together, these issues mean that the usual model-selection criteria do not
  work properly for networks.  We illustrate these challenges, and show one way
  to resolve them, by considering the key network-analysis problem of dividing
  a graph into communities or blocks of nodes with homogeneous patterns of
  links to the rest of the network.  The standard tool for doing this is the
  stochastic block model, under which the probability of a link between two
  nodes is a function solely of the blocks to which they belong.  This imposes
  a homogeneous degree distribution within each block; this can be unrealistic,
  so degree-corrected block models add a parameter for each node, modulating
  its over-all degree.  The choice between ordinary and degree-corrected block
  models matters because they make very different inferences about communities.
  We present the first principled and tractable approach to model selection
  between standard and degree-corrected block models, based on new large-graph
  asymptotics for the distribution of log-likelihood ratios under the
  stochastic block model, finding substantial departures from classical results
  for sparse graphs.  We also develop linear-time approximations for
  log-likelihoods under both the stochastic block model and the
  degree-corrected model, using belief propagation.  Applications to simulated
  and real networks show excellent agreement with our approximations.  Our
  results thus both solve the practical problem of deciding on degree
  correction, and point to a general approach to model selection in network
  analysis.
\end{abstract}


\section{Introduction}

In many networks, nodes divide naturally into modules or communities, where
nodes in the same group connect to the rest of the network in similar
ways. Discovering such communities is an important part of modeling networks
\citep{Porter-Onnela-Mucha-communities}, as community structure offers clues to
the processes which generated the graph, on scales ranging from face-to-face
social interaction \citep{Zachary-karate-club} through social-media
communications \citep{Adamic-Glance-blogosphere} to the organization of food
webs
\citep{Allesina-Pascual-food-webs,Moore-et-al-active-learning-in-networks}.

The stochastic block model
\citep{Fienberg-Wasserman-sociometric-relations,Holland-Lasky-Leinhardt-stochastic-blockmodels,Snijders-Nowicki-blockmodels,Airoldi-Blei-Fienberg-Xing-mixed-membership,Bickel-Chen-on-modularity}
has, deservedly, become one of the most popular generative models for community
detection.  It splits nodes into communities or blocks, within which all nodes
are stochastically equivalent
\citep{Wasserman-Anderson-stochastic-blockmodels}.  That is, the probability of
an edge between any two nodes depends only on which blocks they belong to, and
all edges are independent given the nodes' block memberships.  Block models are
highly flexible, representing assortative, disassortative and satellite
community structures, as well as combinations thereof, in a single generative
framework
\citep{MEJN-assortative-mixing,MEJN-mixing-patterns,Bickel-Chen-on-modularity}.
Their asymptotic properties, including phase transitions in the detectability
of communities, can be determined exactly using tools from statistical physics
\citep{Decelle-et-al-phase-transitions-in-module-detection,Decelle-et-al-asymptotic-analysis-of-stochastic-block-model}
and random graph theory \citep{Mossel-Neeman-Sly-stochastic-block-models}.

Despite this flexibility, stochastic block models impose real restrictions on
networks; notably, the degree distribution within each block is asymptotically
Poisson for large graphs.  This makes the stochastic block model implausible
for many networks, where the degrees within each community are highly
inhomogeneous.  Fitting stochastic block models to such networks tends to split
the high- and low- degree nodes in the same community into distinct blocks; for
instance, dividing both liberal and conservative political blogs into
high-degree ``leaders'' and low-degree ``followers''
\citep{Adamic-Glance-blogosphere,Karrer-MEJN-blockmodels-and-community-structure}.
To avoid this pathology, and allow degree inhomogeneity within blocks, there is
a long history of generative models where the probability of an edge depends on
node attributes as well as their group memberships (e.g.,
\citealt{Morup-Hansen-learning,Reichardt-et-al-micro-macro}).  Here we use the
variant due to \citet{Karrer-MEJN-blockmodels-and-community-structure}, called
the degree-corrected block model\footnote{From a different perspective, the
  famous $p_1$ model of \citet{Holland-Leinhardt-p1}, and the
  \citet{Chung-Lu-model} model, allow each node to have its own expected
  degree, but otherwise treat nodes as homogeneous
  \citep{Rinaldo-Petrovic-Fienberg}.  The degree-corrected block model extends
  these models to allow for systematic variation in linking patterns.}.

We often lack the domain knowledge to choose between the ordinary and the
degree-corrected block model, and so face a model selection problem.  The
standard methods of model selection are largely based on likelihood ratios
(possibly penalized), and we follow that approach here.  Since both the
ordinary and degree-correct block models have many latent variables,
calculating likelihood ratios is itself non-trivial; the likelihood must be
summed over all partitions of nodes into blocks, so (in statistical physics
terms) the log-likelihood is a free energy.  We approximate the log-likelihood
using belief propagation and the Bethe free energy, giving a highly scalable
algorithm that can deal with large sparse networks in nearly linear time.
However, even with the likelihoods in hand, it turns out that the usual
$\chi^2$ theory for likelihood ratios is invalid in our setting, because of a
combination of the sparsity of the data and the high-dimensional nature of the
degree-corrected model.  We derive the correct asymptotics, under regularity
assumptions, recovering the classic results in the limit of large, dense
graphs, but finding that substantial corrections are needed for sparse graphs,
corrections that grow with graph size.  Simulations confirm the validity of our
theory, and we apply our method to both real and synthetic networks.

\section{Poisson Stochastic Block Models}

Let us set the problem on an observed, stochastic graph with $n$ nodes and $m$
edges; we assume edges are undirected, though the directed case is only
notationally more cumbersome.  The graph is represented by its symmetric
adjacency matrix $A$.  We want to split the nodes into $k$ communities, taking
$k$ to be given a priori. (We will address estimating $k$ elsewhere.)

Traditionally, stochastic block models are applied to simple graphs, where each
entry $A_{uv}$ of the adjacency matrix follows a Bernoulli distribution.
Following, e.g., \citet{Karrer-MEJN-blockmodels-and-community-structure}, we
use a multigraph version of the block model, where the $A_{uv}$ are independent
and Poisson-distributed.  (For simplicity, we ignore self-loops.)  In the
sparse network regime we are most interested in, this Poisson mode differs only
negligibly from the original Bernoulli model \citep{Perry-Wolfe-null-models},
but the former is easier to analyze.

\subsection{The Ordinary Stochastic Block Model}

In all stochastic block models, each node $u$ has a latent variable $G_u \in
\{1,\ldots,k\}$ indicating which of the $k$ blocks it belongs to.  The block
assignment is then $G=\{G_u\}$.  The $G_u$ are independent draws from a
multinomial distribution parameterized by $\gamma$, so $\gamma_r = P(G_u = r)$
is the prior probability that a node is in block $r$.  Thus $G_{u} \sim
\text{Multi}(\gamma)$.  After it assigns nodes to blocks, a block model
generates the number of edges $A_{uv}$ between the nodes $u$ and $v$ by making
an independent Poisson draw for each pair.  In the ordinary stochastic block
model, the means of these Poisson draws are specified by the $k \times k$ block
affinity matrix $\omega$, so
\[
A_{uv}|G=g \sim \text{Poi}(\omega_{g_ug_v}) .
\]
The complete-data likelihood (involving $G$ as well as $A$) is
\begin{align}
P(A=a,G=g;\gamma, \omega) 
&= \prod_u{\gamma_{g_u}} \prod_{u<v} \frac{\omega_{g_ug_v}^{a_{uv}} \e^{-\omega_{g_ug_v}}}{a_{uv}!} 
\nonumber \\ 
&= \prod_r{\gamma_r^{n_r}} \prod_{rs} {\omega_{rs}^{m_{rs}/2} \e^{-\frac{1}{2} n_r n_s \omega_{rs}}} \prod_{u<v} \frac{1}{a_{uv}!} . 
\label{eqn:complete-data-likelihood-sbm} 
\end{align}
Here $n_r$ is the number of nodes in block $r$, and $m_{rs}$ the number of
edges connecting block $r$ to block $s$, or twice that number if $r=s$.  The
last product is constant in the parameters, and $1$ for simple graphs, so we
discard it below.  The log-likelihood is then
\begin{equation}
  \label{eq:llh1_dnn}
\like(A=a,G=g;\gamma, \omega)=
  \sum_r{n_r\log{\gamma_r}} + \frac{1}{2} \left( \sum_{rs} {m_{rs}\log{\omega_{rs}}-n_rn_s\omega_{rs}} \right) . 
\end{equation}
Maximizing \eqref{eq:llh1_dnn} over $\gamma$ and $\omega$ gives
\begin{equation}
  \label{eq:mle_sbm}
  \widehat{\gamma}_r = \frac{n_r}{n}
  , \quad
  \widehat{\omega}_{rs} = \frac{m_{rs}}{n_r n_s} .
\end{equation}

Of course, the block assignments $G$ are not observed, but rather are what we
want to infer.  We could try to find $G$ by maximizing \eqref{eq:llh1_dnn} over
$\gamma$, $\omega$ and $g$ jointly; in terms borrowed from statistical physics,
this amounts to finding the ground state $\hat{g}$ that minimizes the energy
$-\like(a,g;\gamma, \omega)$.  When this $\hat{g}$ can be found, it recovers
the correct $g$ exactly if the graph is dense enough
\citep{Bickel-Chen-on-modularity}.  But if we wish to infer the parameters
$\gamma, \omega$, or to perform model selection, we are interested in the total
likelihood of generating the graph $a$ at hand.  This is
\[
P(A=a ; \gamma, \omega) = \sum_g{P(A=a,G=g ; \gamma, \omega)} , 
\]
summing over all $k^n$ possible block assignments.  Again following
the physics lexicon, this is the partition function of the Gibbs distribution
of $G$, and its logarithm is (minus) the free energy.

As is usual with latent variable models, we can infer $\gamma$ and $\omega$
using an EM algorithm \citep{Dempster-Laird-Rubin-introduce-EM}, where the E
step approximates the average over $G$ with respect to the Gibbs distribution,
and the M step estimates $\gamma$ and $\omega$ in order to maximize that
average \citep{Neal-Hinton-view-of-EM}.  One approach to the E step would use a
Monte Carlo Markov Chain 
algorithm to sample $G$ from the Gibbs
distribution.  However, as we review below, in order to determine $\gamma$ and
$\omega$ it suffices to estimate the marginal distributions of $G_u$ of each
$u$, and joint distributions of $(G_u,G_v)$ for each pair of nodes $u, v$
\citep{Beal-Ghahramani-variational-Bayes}.  As we show in \S \ref{sec:BP},
belief propagation efficiently approximates both the log-likelihood
$-\log{P(A=a; \gamma, \omega)}$ and these marginals, and for typical networks
it converges very rapidly.  Other methods of approximating the E step are
certainly possible, and could be used with our model-selection analysis.

\subsection{The Degree-Corrected Block Model}

As discussed above, in the stochastic block model, all nodes in the same block
have the same degree distribution.  Moreover, their degrees are sums of
independent Poisson variables, so this distribution is Poisson.  As a
consequence, the stochastic block model resists putting nodes with very
different degrees in the same block.  This leads to problems with networks
where the degree distributions within blocks are highly skewed.

The degree-corrected model allows for heterogeneity of degree within blocks.
Nodes are assigned to blocks as before, but each node also gets an additional
parameter $\theta_u$, which scales the expected number of edges connecting it
to other nodes.  Thus
\[
A_{uv}|G=g \sim \Poi(\theta_u \theta_v \omega_{g_ug_v}) .
\]
Varying the $\theta_u$ gives any desired expected degree sequence.  Setting
$\theta_u=1$ for all $u$ recovers the stochastic block model.

The likelihood stays the same if we increase $\theta_u$ by some factor $c$ for
all nodes in block $r$, provided we also decrease $\omega_{rs}$ for all $s$ by
the same factor, and decrease $\omega_{rr}$ by $c^2$.  Thus identification
demands a constraint, and here we use the one that forces $\theta_u$ to sum to
the total number of nodes within each block,
\begin{equation}
\label{eq:norm}
\sum_{u:g_u=r}{\theta_u} = n_r .
\end{equation}
The complete-data likelihood of the degree-corrected model is then
\begin{eqnarray}
  \label{eq:lh_DC}
  P(A=a,G=g; \gamma, \omega, \theta)  
  & = &\prod_u{\gamma_{g_u}}\prod_{u<v}{\frac{\left(\theta_u\theta_v\omega_{g_ug_v}\right)^{a_{uv}}}{a_{uv}!} \e^{-\theta_u\theta_v\omega_{g_ug_v}}} \nonumber\\
  & = &\prod_r{\gamma_r^{n_r}} \prod_u{\theta_u^{d_u}
    \prod_{rs} \omega_{rs}^{m_{rs}/2} \e^{-\frac{1}{2} n_r n_s \omega_{rs}}} \prod_{u<v} \frac{1}{a_{uv}!} , 
\end{eqnarray}
where $n_r$ and $m_{rs}$ are as in \eqref{eqn:complete-data-likelihood-sbm}.
Again dropping constants, the log-likelihood is
\begin{multline}
  \label{eq:llh1_DC} 
  \like(A=a,G=g; \gamma, \omega, \theta) \\
  = \sum_r{n_r\log{\gamma_r}} + \sum_u{d_u\log{\theta_u}}+ \frac{1}{2} \left( \sum_{rs}{m_{rs}\log{\omega_{rs}}-n_r n_s \omega_{rs}} \right) . 
\end{multline}
Maximizing \eqref{eq:llh1_DC} yields
\begin{equation}
  \label{eq:mle_DC}
  \widehat{\theta}_u = \frac{d_u}{d_{g_u}}
  , \quad
  \widehat{\gamma}_r=\frac{n_r}{n}
  , \quad 
  \widehat{\omega}_{rs}=\frac{m_{rs}}{d_r d_s} ,
\end{equation}
where
\[
d_r = \frac{1}{n_r} \sum_{u: g_u=r} d_u
\]
is the average degree of the nodes in block $r$.

However, as with the ordinary stochastic block model, we will estimate $\gamma$
and $\omega$ not just for a ground state $\hat{g}$, but using belief
propagation to find the marginal distributions of $G_u$ and $(G_u,G_v)$.

\section{Belief Propagation}
\label{sec:BP}

We referred above to the use of belief propagation for computing
log-likelihoods and marginal distributions of block assignments; for our
purposes, belief propagation is essentially a way of doing the expectation step
of the expectation-maximization algorithm.  Here we describe how belief
propagation works for the degree-corrected block model, extending the treatment
of the ordinary stochastic block model in
\citet{Decelle-et-al-phase-transitions-in-module-detection,Decelle-et-al-asymptotic-analysis-of-stochastic-block-model}.


The key idea \citep{Yedidia-et-al-understanding-belief-propagation} is that
each node $u$ sends a ``message'' to every other node $v$, indicating the
marginal distribution of $G_u$ if $v$ were absent.  We write $\mu^{u \to v}_r$
for the probability that $u$ would be of type $r$ in the absence of $v$.  Then
$\mu^{u \to v}$ gets updated in light of the messages $u$ gets from the other
nodes as follows.  Let
\begin{equation}
  h(\theta_u,\theta_v,\omega_{rs},a_{uv}) = \frac{\left(\theta_u\theta_v\omega_{rs}\right)^{a_{uv}}}{a_{uv}!} 
  \e^{-\theta_u\theta_v\omega_{rs}}
\end{equation}
be the probability that $a_{uv}$ takes its observed value if $G_u=r$ and
$G_v=s$.  Then
\begin{equation}
  \mu^{u \to v}_r = \frac{1}{Z^{u \to v}} \gamma_r \prod_{w \ne u,v}{\sum_{s=1}^k{\mu^{w \to u}_s h(\theta_w,\theta_u,\omega_{rs},a_{wu})}} , 
\end{equation}
where $Z^{u \to v}$ ensures that $\sum_r{\mu^{u \to v}_r} = 1$.  Here, as usual
in belief propagation, we treat the block assignments $G_w$ of the other nodes
as independent, conditioned on $G_u$.

Each node sends messages to every other node, not just to its
neighbors, since non-edges (where $a_{uv}=0$) are also informative about $G_u$
and $G_v$.  Thus we have a Markov random field on a weighted complete graph, as
opposed to just on the network $a$ itself.  However, keeping track of $n^2$
messages is cumbersome.  For sparse networks, we can restore scalability by
noticing that, up to $O(1/n)$ terms, each node $u$ sends the same message to
all of its non-neighbors.  That is, for any $v$ such that $a_{uv}=0$, we have
$\mu_r^{u \to v} = \mu_r^u$ where
\[
  \mu^{u}_r 
  = \frac{1}{Z^u} \gamma_r \prod_{w \ne u}{\sum_{s=1}^k{\mu^{w \to u}_s h(\theta_w,\theta_u,\omega_{rs},a_{wu})}} . 
\]
This simplification reduces the number of messages to $O(n+m)$ where $m$ is the
number of edges.  We can then write
\[
\mu^{u \to v}_r = \frac{1}{Z^{u \to v}} \gamma_r 
\prod_w{\sum_{s=1}^k{\mu^w_s h(\theta_w,\theta_u,\omega_{rs},0)}} 
    \prod_{w \ne v, a_{uw \ne 0}}{ \frac{\sum_{s=1}^k \mu^{w \to u}_s h(\theta_w,\theta_u,\omega_{rs},a_{wu})}
  {\sum_{s=1}^k \mu^w_s h(\theta_w,\theta_u,\omega_{rs},0)}} .
\]

Since the second product depends only on $\theta_u$, we can compute it once for
each distinct degree in the network, and then update the messages for each $u$
in $O(k^2 d_u)$ time.  Thus, for fixed $k$, the total time needed to update all
the messages is $O(m+\ell n)$, where $\ell$ is the number of distinct degrees.
For many families of networks the number of updates necessary to reach a fixed
point is only a constant or $O(\log{n})$, making the entire algorithm quite
scalable (see
\citealt{Decelle-et-al-asymptotic-analysis-of-stochastic-block-model,Decelle-et-al-phase-transitions-in-module-detection} for
details).

The belief-propagation estimate of the joint distribution of $G_u, G_v$ is
\[
b^{uv}_{rs} 
\propto h(\theta_u,\theta_v,\omega_{rs},A_{uv})\mu_r^{u\to v}\mu_s^{v\to u} ,
\]
normalized so that $\sum_{rs} b^{uv}_{rs}=1$.  The maximization step of the
expectation-maximization algorithm sets $\gamma$ and $\omega$ as in
\eqref{eq:mle_DC},
\begin{equation}
  \label{eq:em-update}
  \gamma_r = \frac{\bar{n}_r}{n} = \frac{\sum_u \mu^u_r}{n}
  , \quad
  \omega_{rs} = \frac{\overline{m}_{rs}}{\overline{d}_r \overline{d}_s} 
  = \dfrac{ \sum_{u \ne v: a_{uv}\neq 0} a_{uv} b^{uv}_{rs} }{\sum_u{d_u \mu^u_r} \sum_u{d_u \mu^u_s} } ,
\end{equation}
where $\overline{d}_r$ is the average degree of block $r$ with respect to the
belief-propagation estimates.

Finally, belief propagation also lets us approximate the total log-likelihood,
summed over $G$ but holding the observed graph $a$ fixed.  The Bethe free
energy is the following approximation to the log-likelihood
\citep{Yedidia-et-al-constructing-free-energy}:
\begin{multline}
  \label{eq:bethe}
  \log{P(A=a; \gamma, \omega, \theta)} \\
  \approx \sum_u \log{Z^u} + \frac{1}{2} \sum_{rs} \omega_{rs} \overline{d}_r
  \overline{d}_s - \!\!\! \sum_{u \ne v, a_{uv}\neq 0} \log{\left[\sum_{rs}
      h(\theta_u,\theta_v,\omega_{rs},a_{uv}) \mu_{r}^{u \to v} \mu_{s}^{v \to
        u} \right] } .
\end{multline}

An alternative to belief propagation would be the use of Markov chain Monte
Carlo maximum likelihood, which is often advocated for network modeling
\citep{Hunter-Handcock-inference-in-curved-expo-networks}.  However, the
computational complexity of Monte Carlo maximum likelihood is typically much
worse than that of belief propagation; it does not seem to be practical for
graphs beyond a few hundred nodes.  We reiterate that while we use belief
propagation in our numerical work, our results on model selection in the next
section are quite indifferent as to how the likelihood is computed.

\section{Model Selection}

When the degree distribution is relatively homogeneous within each block (e.g.,
\citealt{Fienberg-Wasserman-sociometric-relations,Holland-Lasky-Leinhardt-stochastic-blockmodels}),
the ordinary stochastic block model is better than the degree-corrected model,
since the extra parameters $\theta_u$ simply lead to over-fitting.  On the
other hand, when degree distributions within blocks are highly heterogeneous,
the degree-corrected model is better. However, without prior knowledge about
the communities, and thus the block degree distributions, we need to use the
data to pick a model, i.e., to do model selection.

It is natural to approach the problem as one of hypothesis testing\footnote{We
  discuss other approaches to model selection in the conclusion.}.  Since the
ordinary stochastic block model is nested within the degree-corrected model,
any given graph $a$ is at least as likely under the latter as under the former.
Moreover, if the ordinary block model really is superior, the degree-corrected
model should converge to it, at least in the limit of large networks.  Our null
model $H_0$ is then the stochastic block model, and the larger, nesting
alternative $H_1$ is the degree-corrected model.  The appropriate test
statistic is the log-likelihood ratio,
\begin{equation}
  \label{eqn:more-explicit-LLR}
  \Lambda(a)  = \log{\frac{\sup_{H_1}{\sum_{g}{P(a,g; \gamma, \omega, \theta)}}}
    {\sup_{H_0}{\sum_g{P(a,g;\gamma, \omega)}}}} ,
\end{equation}
with the $P$ functions defined in \eqref{eqn:complete-data-likelihood-sbm} and
\eqref{eq:lh_DC}.

As usual, we reject the null model in favor of the more elaborate alternative
when $\Lambda$ exceeds some threshold.  This threshold, in turn, is fixed by
our desired error rate, and by the distribution of $\Lambda$ when $A$ is
generated from the null model.  When $n$ is small, the null-model distribution
of $\Lambda$ can be found through parametric bootstrapping \citep[\S
4.2.3.]{Davison-Hinkley-bootstrap}: fitting $H_0$, generating new graphs
$\tilde{A}$ from it, and evaluating $\Lambda(\tilde{A})$.  When $n$ is large, 
however, it is helpful to replace bootstrapping with analytic calculations.

Classically \citep[Theorem 7.125, p. 459]{Schervish-theory-of-stats}, the
large-$n$ null distribution of $2\Lambda$ approaches $\chi^2_\ell$, where
$\ell$ is the number of constraints that must be imposed on $H_1$ to recover
$H_0$.  In this case we have $\ell=n-k$, as we must set all $n$ of the
$\theta_u$ to 1, while our identifiability convention~\eqref{eq:norm}  
already imposed $k$ constraints. 

However, the $\chi^2$ distribution rests on the assumption that the
log-likelihood of both models is well-approximated by a quadratic function in
the vicinity of its maximum, so that the parameter estimates have Gaussian
distributions around the true model \citep{Geyer-on-Le-Cam}.  The most common
grounds for this assumption are central limit theorems for the data, together
with a smooth functional dependence of each parameter estimate on a growing
number of samples, i.e., being in a ``large data limit''.  This assumption 
fails in the present case.  The degree-corrected model has $n$ node-specific
$\theta_u$ parameters.  Dense graphs have an effective sample size of $O(n^2)$,
so even with a growing parameter space the degree-corrected model can pass to
the large data limit.  But in sparse networks, the effective sample size is
only $O(n)$, and so we never get the usual asymptotics no matter how large $n$
grows.

Nevertheless, with some work we are able to compute the mean and variance of
$\Lambda$'s null distribution.  While we recover the classical $\chi^2$
distribution in the limit of dense graphs, there are important corrections when
the average degree of the graph is small, even as $n\to\infty$.  As we shall
see, this has drastic consequences for the appropriate threshold in likelihood
ratio tests.

\subsection{Analysis of the Log Likelihood Ratio}

\begin{figure}
    \begin{center}
      \includegraphics[width=0.5\columnwidth]{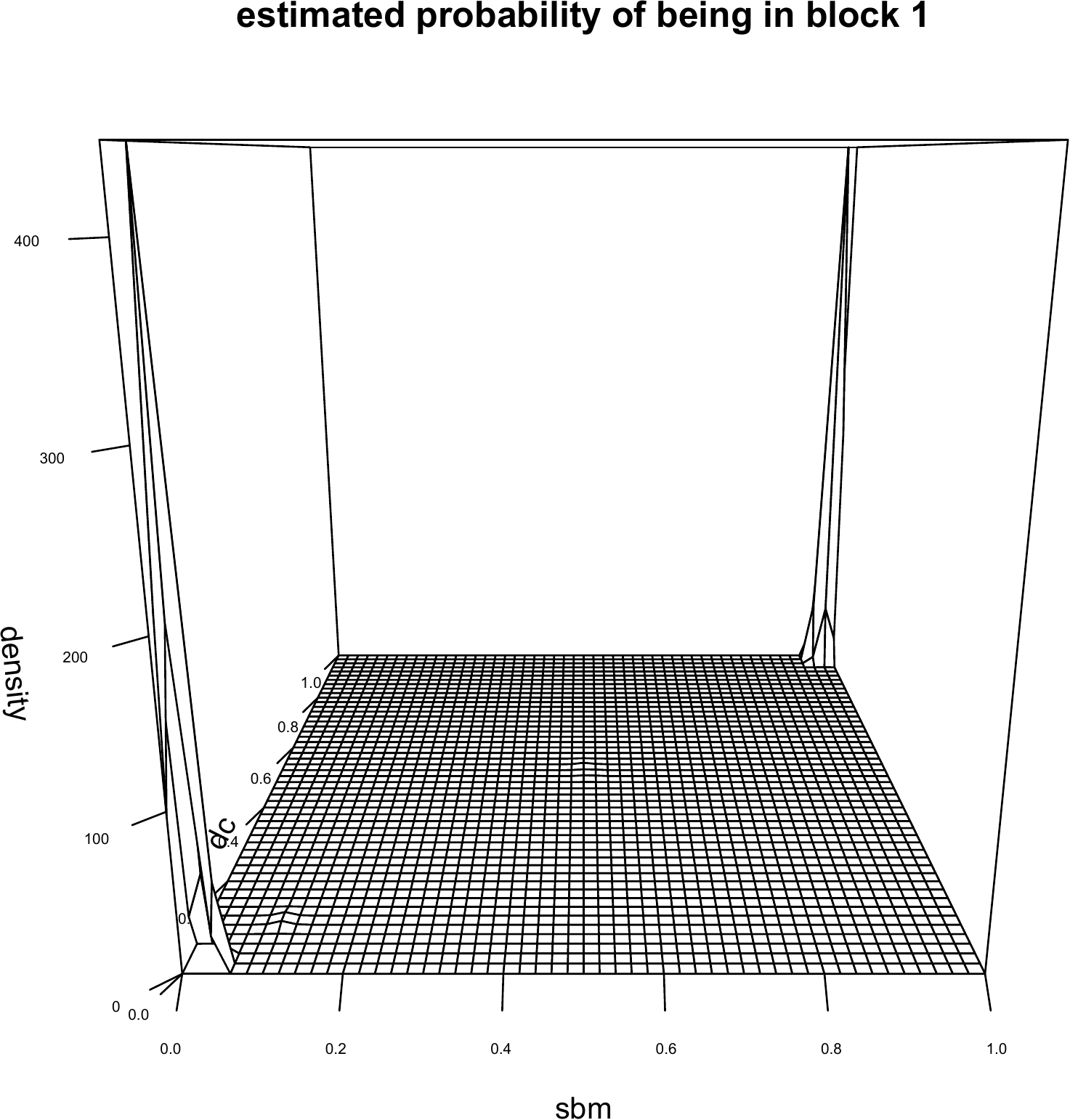}
    \end{center}
    \caption{Joint density of posterior probabilities over block assignments,
      showing that the ordinary and the degree-corrected block models are
      concentrated around the same ground state.  The synthetic network has
      $n={10}^3$, $k=2$ equally-sized blocks ($\gamma_1=\gamma_2=1/2$), average
      degree $\mu_r=11$, and associative structure with
      $\omega_{12}/\omega_{11}=\omega_{21}/\omega_{22}=1/11$.  The $x$ and $y$
      axes are the marginal probabilities of being in block 1 according to
      ordinary and degree-corrected models, respectively.}
  \label{fig:convergence}
\end{figure}

\begin{figure}
  \begin{subfigure}[b]{0.32\textwidth}
    \centering
    \includegraphics[width=\textwidth]{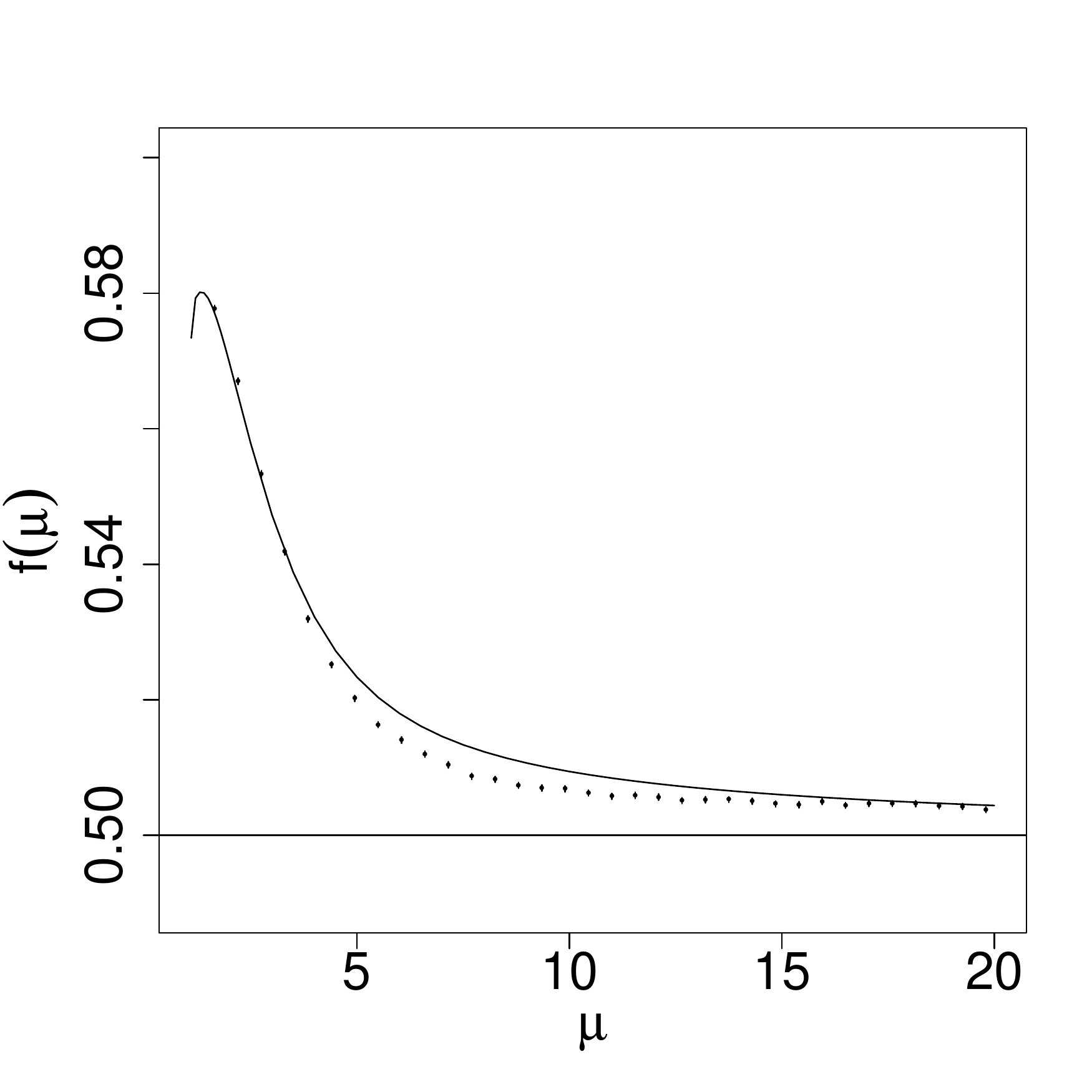}
    \caption{$n^{-1}\Exp{\Lambda}$}
    \label{fig:new-f-of-mu}
  \end{subfigure}
  \begin{subfigure}[b]{0.32\textwidth}
    \centering
    \includegraphics[width=\textwidth]{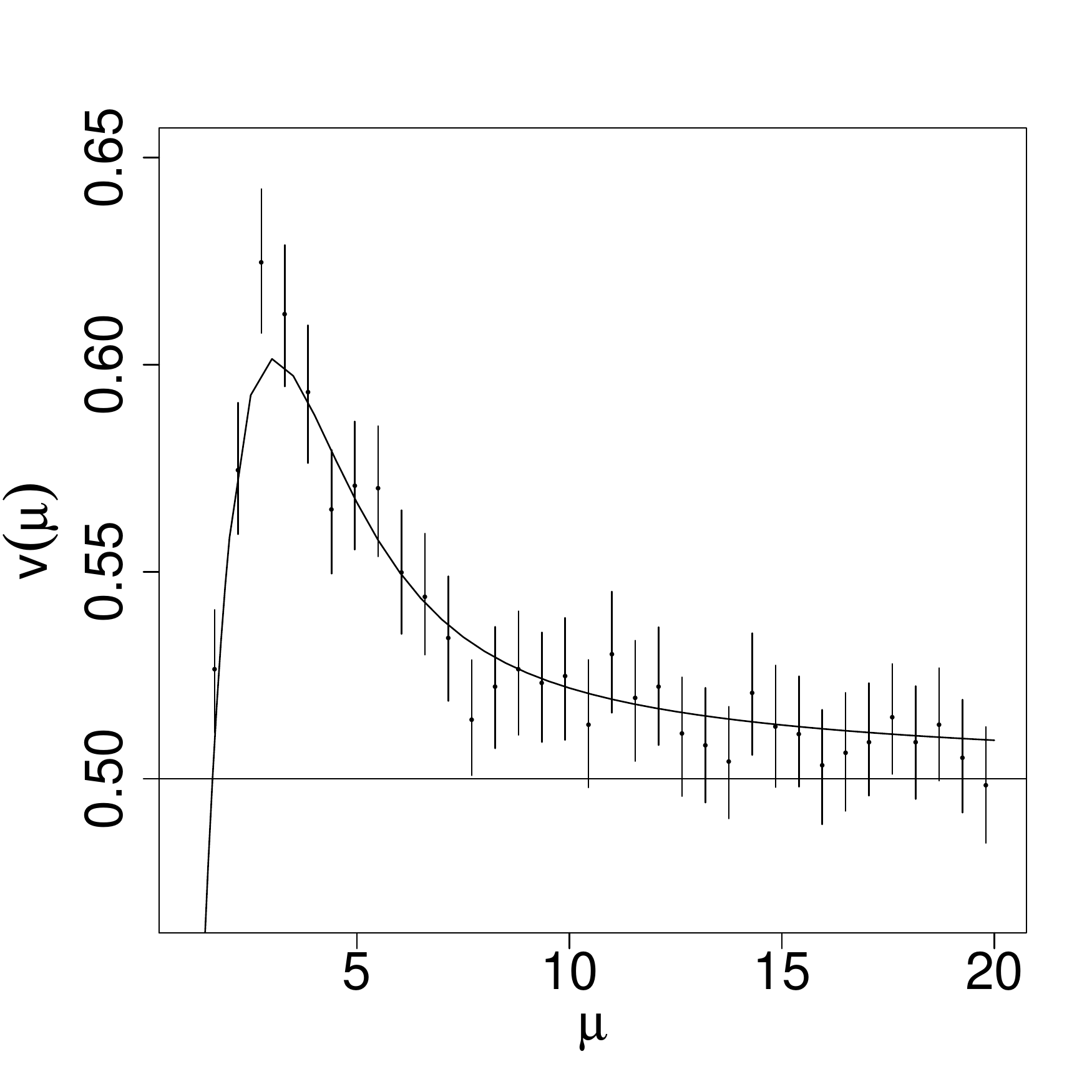}
    \caption{$n^{-1}\Var{\Lambda}$}
    \label{fig:new-v-of-mu}
  \end{subfigure}
  \begin{subfigure}[b]{0.32\textwidth}
    \centering
    \includegraphics[width=\textwidth]{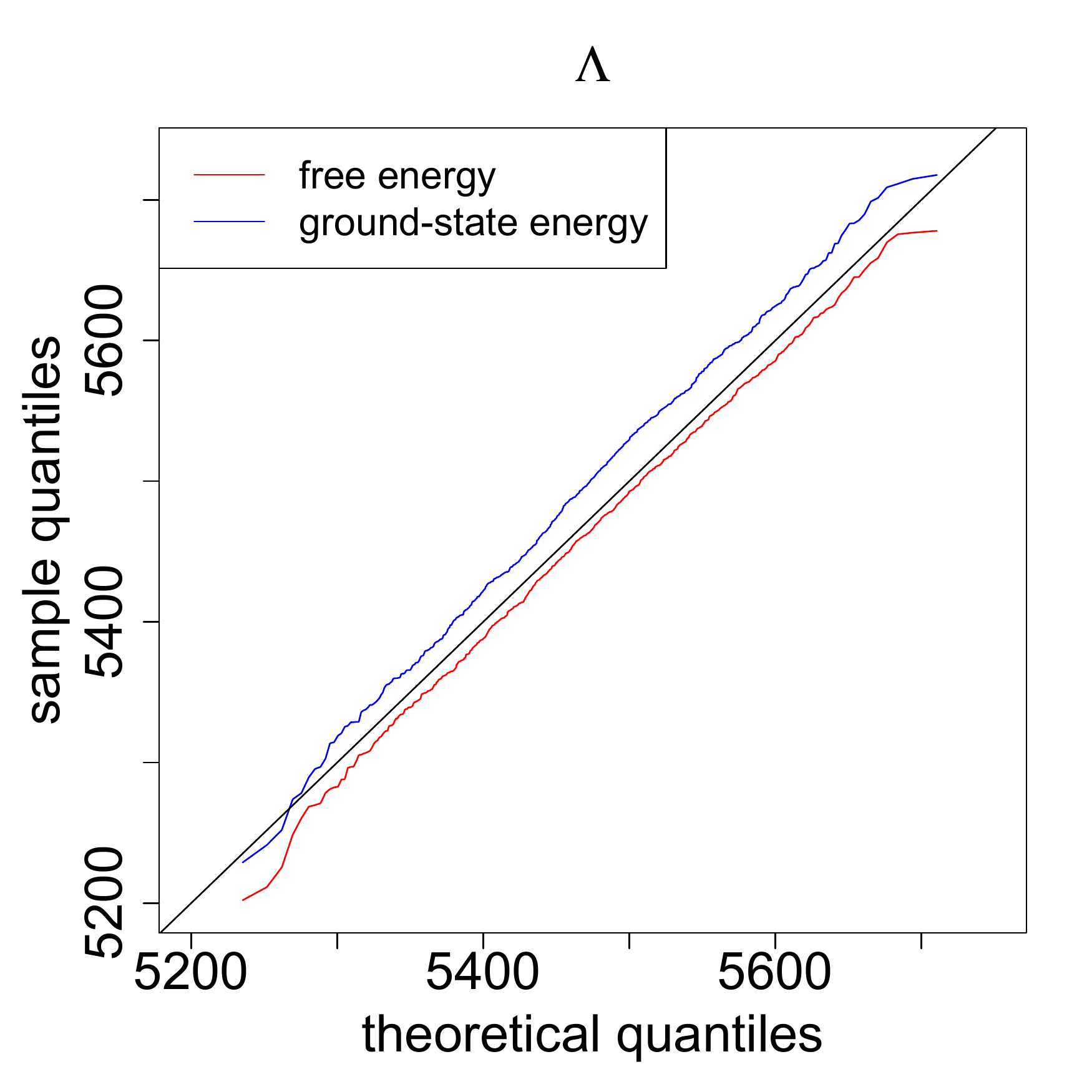}
    \caption{QQ plots for the distribution of $\Lambda$.}
    \label{fig:combined-qq-plots}
  \end{subfigure}
  \caption{Comparison of asymptotic theory to finite-$n$ simulations.  We
    generated networks from the stochastic block model with varying mean degree
    $\mu$ ($n={10}^4$, $k=2$, $\gamma_1=\gamma_2=1/2$, $\omega_{11} =
    \omega_{22}$, and $\omega_{12}/\omega_{11} = \omega_{21}/\omega_{22}
    =0.15$) and computed $\Lambda$ for each graph.  Figures
    \ref{fig:new-f-of-mu} and \ref{fig:new-v-of-mu} show $n^{-1}\Exp{\Lambda}$
    and $n^{-1}\Var{\Lambda}$, comparing 95\% bootstrap confidence intervals
    (over ${10}^3$ replicates) to the asymptotic formulas (respectively
    $f(\mu)$ from \eqref{eqn:expect-log-like-ratio-per-obs} and $v(\mu)$ from
    \eqref{eqn:asymptotic-var-of-delta}).  Figure~\ref{fig:combined-qq-plots}
    compares the distribution of $\Lambda$ from ${10}^4$ replicates, all with
    $\mu=3$, to a Gaussian with the theoretical mean and variance.  (Observe
    that the free energy difference and the ground state energy difference have
    similar distributions.)}
    \label{fig:mu-var}
\label{fig:qq-plots}
\end{figure}

To characterize the null distribution of $\Lambda$, we assume that the
posterior distributions $P(G=g\mid A=a;\gamma,\omega)$ and $P(G=g\mid A=a;
\gamma,\omega,\theta)$ concentrate on the same block assignment $g$.  This is a
major assumption, but it is borne out by our simulations
(Fig.~\ref{fig:convergence} and Fig.~\ref{fig:qq-plots}), and the fact that
under some conditions \citep{Bickel-Chen-on-modularity} the stochastic block
model recovers the underlying block assignment exactly.  Under this assumption,
while the free energy differs from the ground state energy by an entropy term,
the free energy difference between the two models has the same distribution as
the ground state energy difference.  The maximum-likelihood estimates for $H_0$
and $H_1$ are then \eqref{eq:mle_sbm} and \eqref{eq:mle_DC} respectively.
Substituting these into \eqref{eqn:more-explicit-LLR}, most of the terms
cancel, giving $\Lambda$ the form of a Kullback-Leibler divergence,

\begin{align}
\label{eq:lambda}
\Lambda &\approx \log{\frac{\sup_{H_1}{{\prod_u{\theta_u^{d_u}}\prod_r{
          q_r^{n_r}}\prod_{rs}{\omega_{rs}^{m_{rs}/2} \e^{-\frac{1}{2} n_r n_s
            \omega_{rs}}}}}} {\sup_{H_0} \prod_r{
      q_r^{n_r}}\prod_{rs}{\omega_{rs}^{m_{rs}/2} \e^{-\frac{1}{2} n_r n_s
        \omega_{rs}}}}}
\nonumber\\
&= \log{\sup_{H_1} \prod_u \theta_u^{d_u}} = \log{ \prod_u \left(
    \frac{d_u}{d_{g_u}} \right)^{d_u}} = \sum_u{d_u \log{\frac{d_u}{d_{g_u}}}}
.
\end{align}
where we applied~\eqref{eq:mle_DC}.  Here $d_r$ is the empirical mean degree of
block $r$, not the expected degree $\mu_r = \sum_s{\gamma_s \omega_{rs}}$ of
the stochastic block model.


Given \eqref{eq:lambda}, the distribution of $\Lambda$ follows from the
distributions of the nodes' degrees; under the null model, all the $D_u$ in
block $r$ are independent $\sim \Poi(\mu_r)$.  (This assumption is sound in the
limit $n \to \infty$, since the correlations between node degrees are
$O(1/n)$.)  Using this, we can compute the expectation and variance of
$\Lambda$ analytically (see Appendix), showing that $\Lambda$ departs from
classical $\chi^2$ asymptotics, as well as revealing the limits where those
results apply.  Specifically,
\begin{equation}
  \label{eq:exp-lambda}
  \Exp{\Lambda} = \sum_r{n_rf(\mu_r) - f(n_r \mu_r)}
\end{equation}
where, if $D \sim \Poi(\mu)$,
\begin{equation}
  \label{eqn:expect-log-like-ratio-per-obs} 
  f(\mu) = \Exp{D\log{D}} - \mu\log{\mu} . 
\end{equation}
For dense graphs, where $\mu \to \infty$, both $f(\mu)$ and $f(n\mu)$ approach
$1/2$, and \eqref{eq:exp-lambda} gives $\Exp{\Lambda} = (n-k)/2$ just as in the
standard $\chi^2$ analysis.  However, when $\mu$ is small, $f(\mu)$ differs
noticeably from $1/2$.

The variance of $\Lambda$ is somewhat more complicated.  The limiting variance
per node is
\begin{equation}
  \label{eq:var-lambda}
  \lim_{n \to \infty} \frac{1}{n} \Var{\Lambda} = \sum_r{\gamma_r v(\mu_r)} , 
\end{equation}
where, again taking $D \sim \Poi(\mu)$,
\begin{equation}
  \label{eqn:asymptotic-var-of-delta}
  v(\mu) = \mu(1+\log{\mu})^2 + \Var{D\log{D}} - 2(1+\log{\mu})\Cov{D,D\log{D}} .
\end{equation}
Since the variance of $\chi^2_\ell$ is $2\ell$, $\chi^2$ asymptotics would
predict $(1/n) \Var{\Lambda} = 1/2$.  Indeed $v(\mu)$ approaches $1/2$ as $\mu
\to \infty$, but like $f(\mu)$ it differs substantially from $1/2$ for small
$\mu$.  Figure~\ref{fig:mu-var} plots $f(\mu)$ and $v(\mu)$ for $1 \le \mu \le
10$.

\begin{figure}
  \begin{subfigure}[b]{0.48\textwidth}
    \centering
    \includegraphics[width=\textwidth]{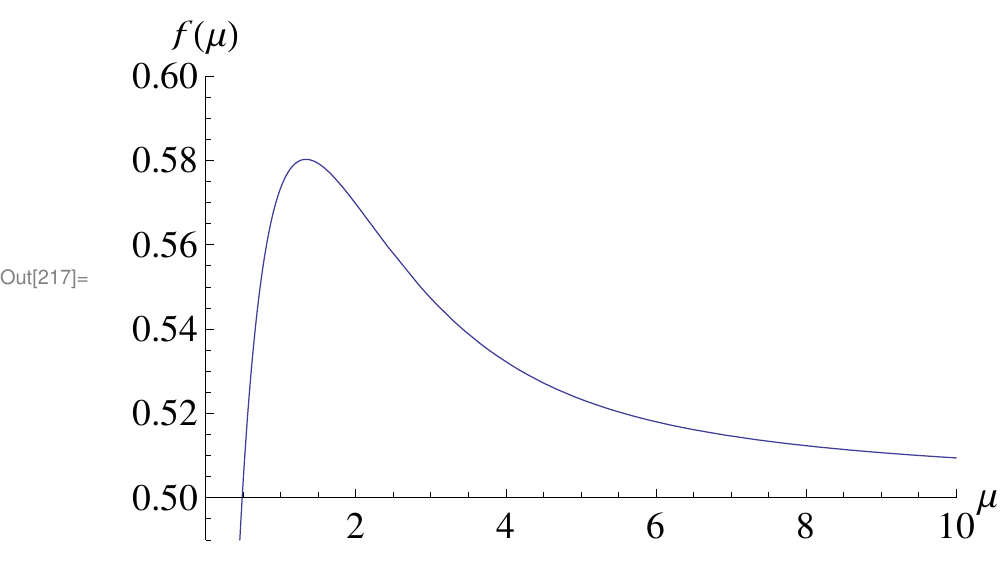}
    \caption{$f(\mu)$}
  \end{subfigure}
  \begin{subfigure}[b]{0.48\textwidth}
    \centering
    \includegraphics[width=\textwidth]{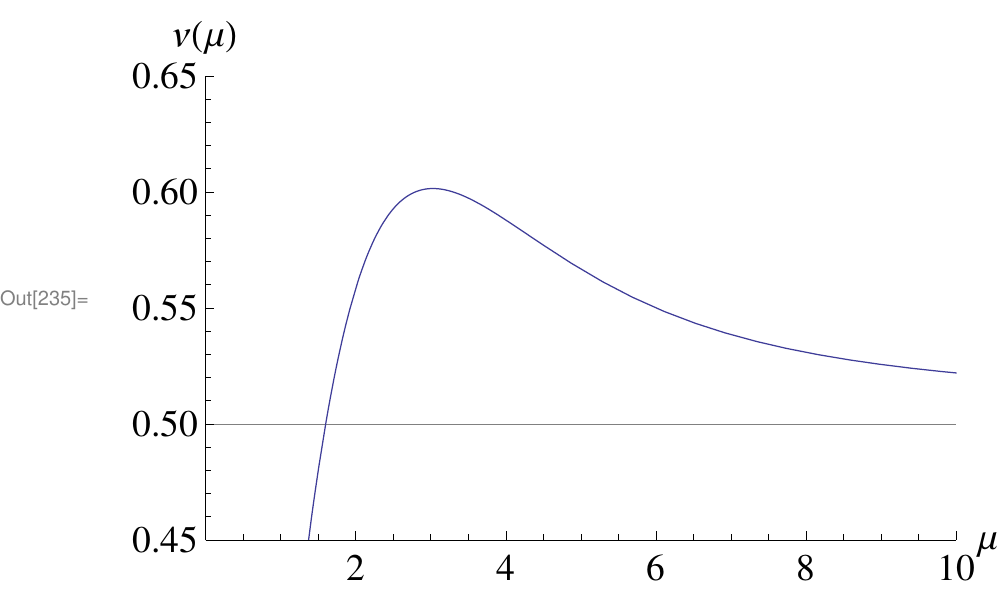}
    \caption{$v(\mu)$}
    \label{fig:var}
  \end{subfigure}
  \caption{(a) Asymptotic limit of $n^{-1}\Exp{\Lambda}$, the $f(\mu)$ of
    \eqref{eq:f}.  (b) Asymptotic limit of $n^{-1}\Var{\Lambda}$, from
    \eqref{eq:varOn}.  Figure~\ref{fig:qq-plots} compares these to
    simulations.}
  \label{fig:f}
\end{figure}

Figure \ref{fig:mu-var} shows that, for networks simulated from the stochastic
block model, the mean and variance of $\Lambda$ are very well fit by our
formulas.  We have not attempted to compute higher moments of $\Lambda$.
However, if we assume that $D_u$ are independent, then the simplest form of the
central limit theorem applies, and $n^{-1}\Lambda$ will approach a Gaussian
distribution as $n\to\infty$.  Quantile plots from the same simulations
(Fig.~\ref{fig:qq-plots}(c)) show that a Gaussian with mean and variance from
\eqref{eq:exp-lambda} and \eqref{eq:var-lambda} is indeed a good fit.
Moreover, the free energy difference and the ground state energy difference
have similar distributions, as implied by our assumption that both Gibbs
distributions are concentrated around the ground state.  Interestingly, in
Fig.~\ref{fig:qq-plots}(c), the degree is low enough that this concentration
must be imperfect, but our theory still holds remarkably well.  For ease of
illustration, we assume that $\gamma_r = 1/k$ and $\mu_r$ are the same for all
$r$.

Fundamentally, $\Lambda$ does not follow the usual $\chi^2$ distribution
because the $\theta$ parameters are in a high-dimensional regime.  For each
$\theta_u$, we really have only one relevant observation, the node degree
$D_u$.  If $\theta_u$ is large, then the Poisson distribution of $D_u$ is
well-approximated by a Gaussian, as is the sampling distribution of
$\theta_u$'s maximum likelihood estimate, so that the usual $\chi^2$ analysis
applies.  In a sparse graph, however, all the Poisson distributions have small
expected values and are highly non-Gaussian, as are the maximum likelihood
estimates \citep{Zhu-Yan-Moore-inhomogeneous-degree}.  Said differently, the
degree-corrected model has $O(n)$ more parameters than the null model.  In the
dense-graph case, there are $O(n^2)$ observations, at least $O(n)$ of which are
informative about each of these extra parameters.  For sparse graphs, however,
there are really only $O(n)$ observations, and only $O(1)$ of them are
informative about each $\theta_u$, so the ordinary large-$n$ asymptotics cannot
apply to them.  As we have seen, the expected increase in likelihood from
adding the $\theta$ parameters is larger than $\chi^2$ theory predicts, as are
the fluctuations in this increase in likelihood.

This reasoning elaborates on a point made long ago by
\citet{Fienberg-Wasserman-comment-on-Holland-Leinhardt} regarding hypothesis
testing in the $p_1$ model, where each node has two node-specific parameters
(for in- and out- degree); our calculations of $f(\mu)$ and $v(\mu)$ above, and
especially of how and why they differ from $1/2$, go some way towards meeting
Fienberg and Wasserman's call for appropriate asymptotics for large-but-sparse
graphs.

\begin{figure}
  \begin{center}
    \includegraphics[width=0.5\columnwidth]{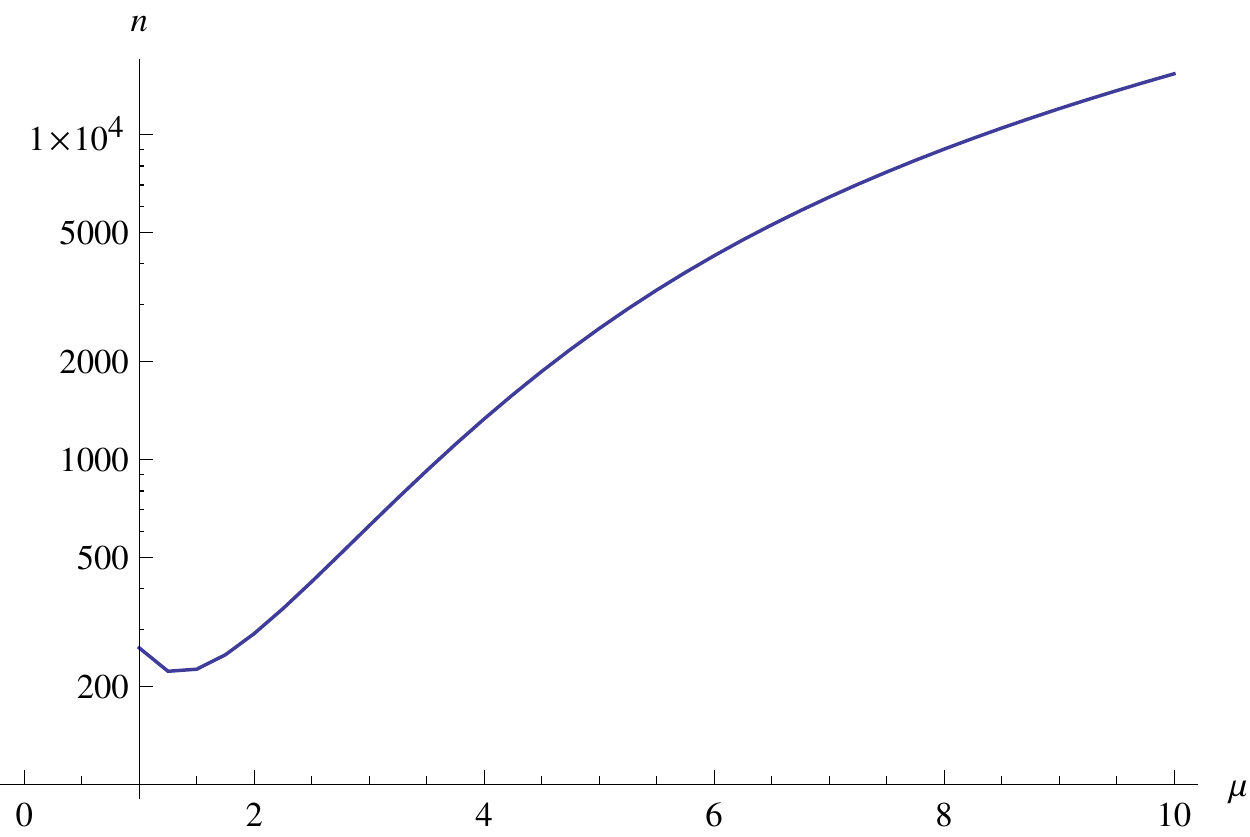}
  \end{center}
  \caption{The network size $n$, as a function of the average degree $\mu$,
    above which a $\chi^2$ test with a nominal type I error rate of $0.05$ has
    an actual type I error rate $\geq 0.95$.  Here we assume the asymptotic
    analysis of \eqref{eq:exp-lambda}--\eqref{eqn:asymptotic-var-of-delta} for
    the mean and variance of the likelihood ratio.}
  \label{fig:errorScaling}
\end{figure}

Ignoring these phenomena and using a $\chi^2$ test inflates the type I error
rate ($\alpha$), eventually rejecting the stochastic block model for almost all
graphs which it generates.  Indeed, since the $\chi^2$ distribution is tightly
peaked around $0.5n$, this inflation of $\alpha$ gets worse as $n$ gets bigger.
For instance, when $\mu=5$, a $\chi^2$ test with nominal $\alpha=0.05$ passes
has a true $\alpha \geq 0.95$ once $n \approx 3000$, while for $\mu=3$, this
happens once $n\approx 700$ (Fig.  \ref{fig:errorScaling}). In essence, the
$\chi^2$ test underestimates the amount of degree inhomogeneity we would get
simply from noise, incorrectly concluding that the inhomogeneity must come from
underlying properties of the nodes.

\section{Results on real networks}

We have derived the theoretical null distribution of $\Lambda$, and backed up
our calculations with simulations.  We now apply our theory to two examples,
considering networks studied in
\citet{Karrer-MEJN-blockmodels-and-community-structure}.

The first is a social network consisting of $34$ members of a karate club,
where undirected edges represent friendships \citep{Zachary-karate-club}. The
network is made up of two assortative blocks, each with one high-degree hub
(respectively the instructor and the club president) and many low-degree
peripheral nodes.  \citet{Karrer-MEJN-blockmodels-and-community-structure}
compared the performance of the ordinary and the degree-corrected block models
on this network, and heavily favored degree correction, because the former
leads to division into communities agreeing with ethnographic observations.

\begin{figure}
  \begin{subfigure}[b]{0.48\textwidth}
    \centering
    \includegraphics[width=\textwidth]{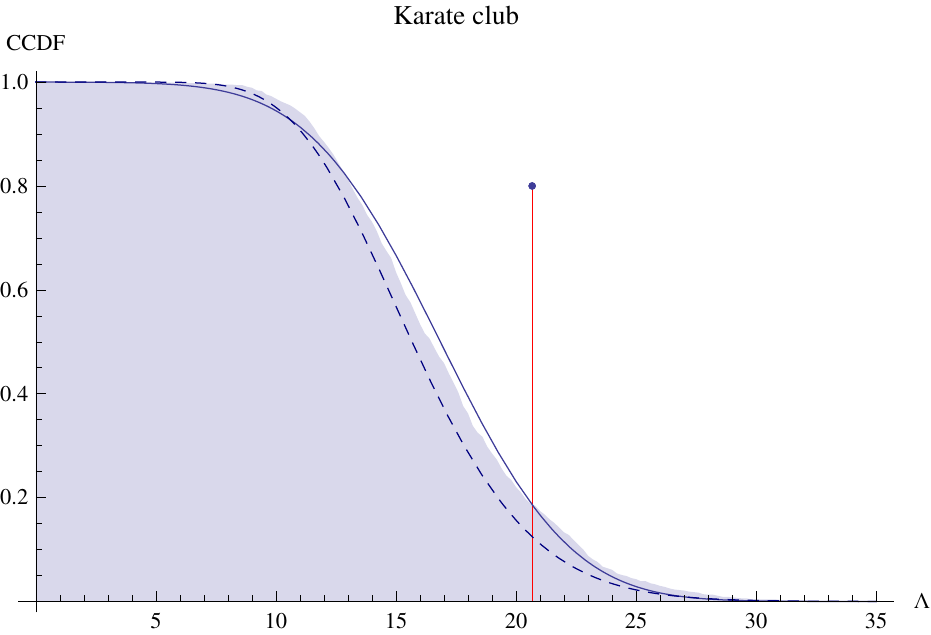}
  \end{subfigure}
  \begin{subfigure}[b]{0.48\textwidth}
    \centering
    \includegraphics[width=\textwidth]{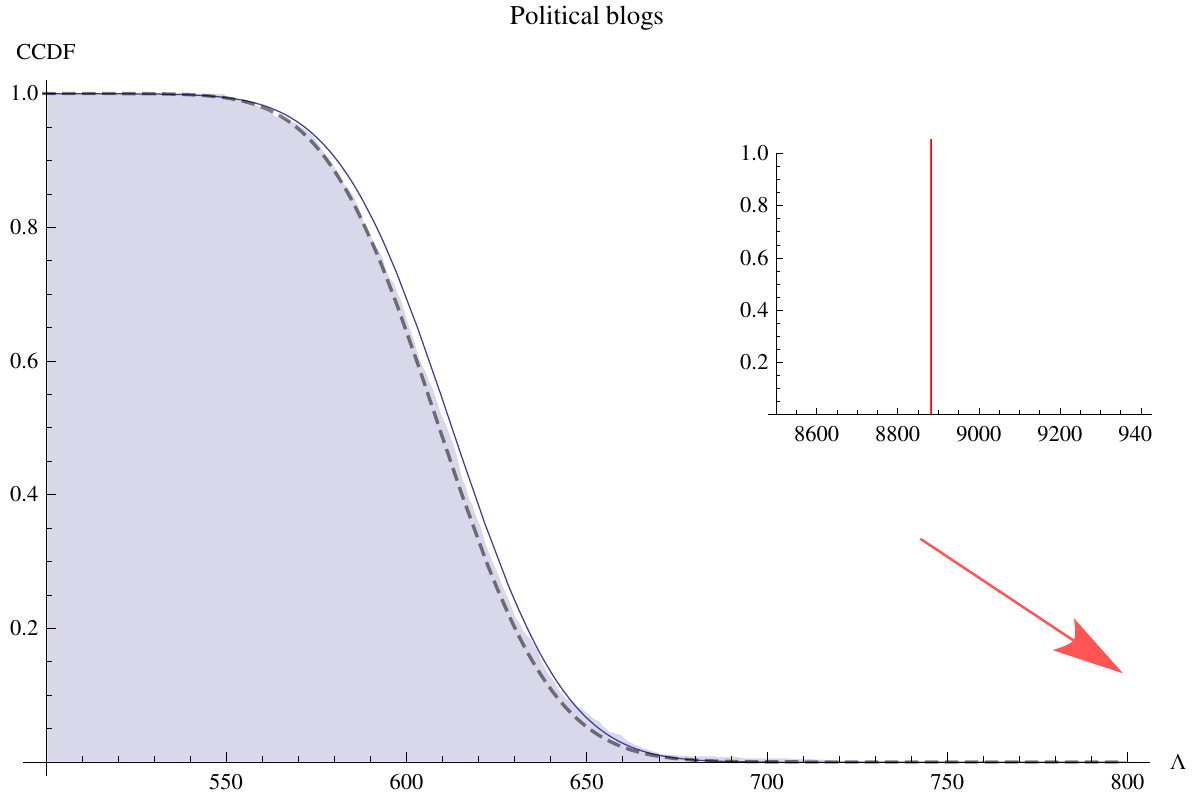}
  \end{subfigure}
  \caption{Hypothesis testing with real networks; both panels show the
    complementary cumulative distribution function of the log-likelihood ratio
    $\Lambda$ for testing for degree correction.  (a): Zachary's karate club
    \citep{Zachary-karate-club} ($n=34$).  The distribution found by parametric
    bootstrapping (shaded) fits reasonably well to a Gaussian (curve) with our
    theoretical mean and variance.  The observed $\Lambda=20.7$ (marked with
    the red line) has $p$-values of $0.186$ and $0.187$ according to the
    bootstrap and theoretical distributions respectively, whereas the $\chi^2$
    test (dashed) has a $p$-value of $0.125$.  (b): A network of political
    blogs \citep{Adamic-Glance-blogosphere} ($n=1222$).  The bootstrap
    distribution (shaded) is very well fit by our theoretical Gaussian (curve)
    as well as the $\chi^2$ test (dashed). The actual log-likelihood ratio is
    so far in the tail that its $p$-value is effectively zero (see inset).
    Thus for the blog network, we can decisively reject the ordinary block
    model in favor of the degree-corrected model, while for the karate club,
    the evidence is less clear.  }
  \label{fig:realworld}
\end{figure}

While a classic data set for network modeling, the karate club has both low
degree and very small $n$.  If we nonetheless use parametric bootstrapping to
find the null distribution of $\Lambda$, we see that it fits a Gaussian with
our predicted mean and variance reasonably well (Fig.~\ref{fig:realworld}(a)).
The observed $\Lambda=20.7$ has a $p$-value of $0.187$ according to the
bootstrap, and $0.186$ according to our Gaussian asymptotics.  Thus a prudent
analyst would think twice before embracing the $n$ additional degree-correction
parameters.  Indeed, using active learning,
\citet{Moore-et-al-active-learning-in-networks} found that the stochastic block
model labels most of the nodes correctly if the instructor and president are
forced into different blocks.  This implies that the degree inhomogeneity is
mild, and that only a handful of nodes are responsible for the better
performance of the degree-corrected model.  

Note that if we apply standard $\chi^2$ testing to the karate club, we obtain a lower $p$-value of $0.125$.  
As in Fig.~\ref{fig:errorScaling}, $\chi^2$ testing underestimates the extent to which an inhomogeneous 
degree distribution can result simply from noise, causing it to reject the null model more confidently than it should.

The second example is a network of political blogs in the US assembled by
\citet{Adamic-Glance-blogosphere}.  As in
\citet{Karrer-MEJN-blockmodels-and-community-structure}, we focus on the giant
component, which contains $1222$ blogs with $19087$ links between them.  The
blogs have known political leanings, and were labeled as either liberal or
conservative.  The network is politically assortative, with highly right-skewed
degree distributions within each block, so degree correction greatly assists in
recovering political divisions, as observed by
\citet{Karrer-MEJN-blockmodels-and-community-structure}.  This time around, our
hypothesis testing procedure completely agrees with their choice of model.  As
shown in Fig.~\ref{fig:realworld}(b), the bootstrap distribution of $\Lambda$
is very well fit by a Gaussian with our theoretical prediction of the mean and
variance.  The observed log-likelihood ratio $\Lambda=8883$ is $330$ standard
deviations above the mean.  It is essentially impossible to produce such
extreme results through mere fluctuations under the null model.  Thus, for this
network, introducing $n$ extra parameters to capture the degree heterogeneity
is fully justified.

The blog network shows several advantages of our theoretical approach over just
using bootstrapping.  As with many other real networks, $n$ is large enough
that bootstrapping is quite slow, but for the same reason the Gaussian
approximation for $\Lambda$ is fairly tight.

\section{Conclusion}

Deciding between ordinary and degree-corrected stochastic block models for
sparse graphs presents a difficult hypothesis testing problem.  The
distribution of the log-likelihood ratio $\Lambda$ does not follow the classic
$\chi^2$ theory, because the nuisance parameter $\theta$, only present in the
alternative, is in a high-dimensional regime.  We have nonetheless derived
$\Lambda$'s mean and variance in the limit of large, sparse graphs, where node
degrees become independent and Poisson.  Simulations confirm the accuracy of
our theory for moderate $n$, and we applied it to two real networks.

Beyond hypothesis testing, two standard approaches to model selection are
information criteria and cross-validation.  While we have not directly dealt
with the former, the derivations of such popular criteria as AIC or DIC use
exactly the same asymptotics as the $\chi^2$ test \citep[ch.\
2]{Claeskens-Hjort-model-selection}; these tools will break down for the same
reasons $\chi^2$ theory fails.  As for cross-validation, standard practice in
machine learning suggests using multi-fold cross-validation, but the global
dependence of network data means there is (as yet) no good way to split a graph
into training and testing sets.  Predicting missing links or tagging false
positives are popular forms of leave-$k$-out cross-validation in the network
literature
\citep{Clauset-Moore-MEJN-hierarchical-structure,Guimera-and-Sales-Pardo-missing-and-spurious},
but leave-$k$-out does not converge on the true model even for independent and
identically-distributed data \citep[\S 2.9]{Claeskens-Hjort-model-selection}.
Thus, while our results apply directly only to the specific problem of testing
the need for degree correction, they open the way to more general approaches to
model selection and hypothesis testing in a wide range of network problems.

\appendix

\section{Behavior of $\Lambda$ under the null hypothesis}
\label{app:lambda}

For simplicity we focus on one block with expected degree $\mu$.  Independence
between blocks will then recover the expressions \eqref{eq:exp-lambda} and
\eqref{eq:var-lambda} where the mean and variance of $\Lambda$ is a weighted
sum over blocks.  We have
\begin{equation}
\Lambda 
= \sum_{i=1}^n {D_i \log{\frac{D_i}{\Dbar}}} 
= \sum_i{D_i \log{D_i}} - \left( \sum_i{D_i} \right) \log{\left( \sum_i{D_i} \right)} + \left( \sum_i{D_i} \right) \log{n} ,  
\label{eq:delta-log-like}
\end{equation}
where $\Dbar = (1/n) \sum_i{D_i}$ is the empirical mean degree.  We wish to
compute the mean and expectation of $\Lambda$ if the data is generated by the
null model.

If $D \sim \Poi(\mu)$, let $f(\mu)$ denote the difference between the
expectation of $D \log{D}$ and its most probable value $\mu \log{\mu}$:
\begin{equation}
  \label{eq:f}
  f(\mu) = \left( \sum_{d=1}^\infty \frac{\e^{-\mu} \mu^d}{d!} d \log{d} \right) - \mu \log \mu  . 
\end{equation}
Assume that the $D_i$ are independent and $\sim \Poi(\mu)$; this is reasonable
in a large sparse graph, since the correlations between degrees of different
nodes is $O(1/n)$.  Then $n\Dbar \sim \Poi(n\mu)$, and
\eqref{eq:delta-log-like} gives
\begin{equation}
  \label{eq:exp-delta-log-like}
  \Exp{\Lambda} = n f(\mu) - f(n \mu)  . 
\end{equation}
To grasp what this implies, begin by observing that $f(\mu)$ converges to $1/2$
when $\mu$ is large.  Thus in the limit of large $n$, $\Exp{\Lambda} = n f(\mu)
- \frac{1}{2} $.  When $\mu$ is large, this gives $\Exp{\Lambda} = (n-1)/2$,
just as $\chi^2$ theory suggests.  However, as Fig.~\ref{fig:f} shows, $f(\mu)$
deviates noticeably from $1/2$ for finite $\mu$.  We can obtain the leading
corrections as a power series in $1/\mu$ by approximating \eqref{eq:f} with the
Taylor series of $d \log d$ around $d=\mu$, giving
\[
f(\mu) = \frac{1}{2} + \frac{1}{12 \mu} + \frac{1}{12 \mu^2} + O(1/\mu^3) .
\]

Computing the variance is harder.  It will be convenient to define several
functions.  If $D \sim \Poi(\mu)$, let $\phi(\mu)$ denote the variance of $D
\log{D}$:
\begin{equation}
  \phi(\mu) = \Var{D \log{D}} 
  = \sum_{d=0}^\infty \frac{\e^{-\mu} \mu^d}{d!} (d \log d)^2 - \left( f(\mu) + \mu \log{\mu} \right)^2 . 
\label{eq:v}
\end{equation}
We will also use
\begin{equation}
  c(\mu) = \Cov{D, D \log{D} }
  = \sum_{d=1}^\infty \frac{\e^{-\mu} \mu^d}{d!} d^2 \log d - \mu \left( f(\mu) + \mu \log \mu \right) . 
  \label{eq:c}
\end{equation}
Finally, let $\psi \ge \mu$, and let $D$ and $U$ be independent and Poisson
with mean $\mu$ and $\psi-\mu$ respectively.  Then let
\begin{eqnarray}
  r(\mu,\psi) &= & \Cov{D \log D, (D+U) \log (D+U)}    \nonumber\\
  &= & \Exp{(D \log D)((D+U) \log (D+U))} - \Exp{D \log D} \Exp{(D+U) \log (D+U)} 
  \nonumber \\
  &= & \Exp{(D \log D)((D+U) \log (D+U))} - \left( f(\mu)+\mu \log \mu \right) \left( f(\psi)+ \psi \log{\psi} \right) ,   \label{eq:r}
\end{eqnarray}
where we used the fact that $D+U \sim \Poi(\psi)$.

Again assuming that the $D_i$ are independent, we have the following terms and
cross-terms for the variance of \eqref{eq:delta-log-like} :
\begin{align*}
  \Var{\sum_i{D_i \log{D_i}}}
  &= n \phi(\mu) \\
  \Var{\left( n\Dbar \right) \log{\left( n\Dbar \right)}}
  &= \phi(n \mu) \\
  \Var{n\Dbar}
  &= n \mu \\
  \Cov{\sum_i{D_i \log{D_i}}, \left( n\Dbar \right) \log{\left( n\Dbar
      \right)}}
  &= n r(\mu, n\mu) \\
  \Cov{\sum_i{D_i \log{D_i}}, n\Dbar}
  &= nc(\mu) \\
  \Cov{\left( n\Dbar \right) \log{n\Dbar} , n\Dbar } &= c(n \mu)
\end{align*}
Putting this all together, we have
\begin{equation}
  \label{eq:var}
  \Var{\Lambda} 
  = n \phi(\mu) + \phi(n\mu) + n \mu \log^2{n} - 2n r(\mu, n\mu) + 2 \left( nc(\mu) - c(n\mu) \right) \log{n}  . 
\end{equation}

For large $\mu$, Taylor-expanding the summands of \eqref{eq:v} and \eqref{eq:c}
yields
\begin{align*}
  \phi(\mu) &= \mu \log^2{\mu} + 2\mu \log{\mu} + \mu + \frac{1}{2} + O\left( \frac{\log \mu}{\mu} \right) \\
  c(\mu) &= \mu \log{\mu} + \mu + O(1/\mu) .
\end{align*}
Also, when $\psi \gg \mu$ and $\mu=O(1)$, using $\log{(D+U)} \approx \log U +
D/U$ lets us simplify \eqref{eq:r}, giving
\begin{align*}
r(\mu,\lambda) 
&= \Exp{D^2 \log{D}} (1 + \log \lambda) \\
&+ \Exp{D \log{D}} \Exp{U \log U} \\ 
&- \Exp{D \log{D}} \Exp{(D+U) \log{D+U}}) + O(1/\lambda) .
\end{align*}

In particular, setting $\psi = n \mu$ gives
\[
r(\mu, n\mu) = c(\mu) (1+\log n\mu) + O(1/n) .
\]

Finally, keeping $O(n)$ terms in \eqref{eq:var} and defining $v(\mu)$ as in
\eqref{eq:var-lambda} gives
\begin{equation}
\label{eq:varOn}
v(\mu)  = \lim_{n\to\infty} \frac{1}{n} \Var{\Lambda} = \phi(\mu) + \mu (1+\log \mu)^2 - 2 c(\mu) (1+\log \mu) . 
\end{equation}
Using the definitions of $\phi$ and $c$, we can write this more explicitly as
\begin{equation}
\label{eq:varOn2}
v(\mu) = \mu(1+\log \mu)^2 + \Var{D \log{D}} -2 (1+\log \mu) \Cov{D, D \log{D}} ,
\end{equation}
where $D \sim \Poi(\mu)$.  We plot this function in Fig.~\ref{fig:var}.  It
converges to $1/2$ in the limit of large $\mu$, but it is significantly larger
for finite $\mu$.

\bibliography{locusts}
\bibliographystyle{biometrika}

\end{document}